\documentclass{amsart}
 \usepackage{amsbsy,amssymb,amscd,amsfonts,latexsym,amstext,delarray,
 amsmath,graphicx}
 \usepackage[dvips]{epsfig}
\usepackage{hyperref}

\def\cutint{{\int \!\!\!\!\!\! -}}

\newtheorem{thm}{Theorem}[section]
\newtheorem{prop}[thm]{Proposition}
\newtheorem{cor}[thm]{Corollary}
\newtheorem{lem}[thm]{Lemma}

\newtheorem{defn}[thm]{Definition}
\newtheorem{rem}[thm]{Remark}

\def\qqq{\,,\quad \forall}

\def\Dom{{\rm Dom}}
\def\End{{\rm End}}

\def\SU{{\rm SU}}

\def\Sp{{\rm Spec}}

\def\Tr{{\rm Tr}}
\def\tr{{\rm tr}}

\def\C{{\mathbb C}}

\def\N{{\mathbb N}}
\def\P{{\mathbb P}}

\def\R{{\mathbb R}}
\def\Z{{\mathbb Z}}

\def\Tr{{\rm Tr}}
\def\tr{{\rm tr}}

\def\cA{{\mathcal A}}

\def\cH{{\mathcal H}}

\def\cL{{\mathcal L}}

\def\cS{{\mathcal S}}

\def\cU{{\mathcal U}}

\newcommand{\ie}{{\it i.e.\/}\ }
\newcommand{\alm}{{\it a.e.\/}\ }
\newcommand{\eg}{{\it e.g.\/}\ }
\newcommand{\cf}{{\it cf.\/}\ }

\def\sin{{{\rm sin}}}
\def\cos{{{\rm cos}}}

\def\dim{{\mbox{dim}}}
\def\ker{{\mbox{Ker}}}

\def\End{{\mbox{End}}}

\def\Ss{{\bf S}}

\def\gra{{\bf Gr}}
\def\her{{\bf H}}
\def\per{{\bf P}}

\newcommand{\nil}[1]{}

\def\axiom{condition }
\def\axioms{conditions }

\title
{A unitary invariant in Riemannian Geometry}
\author[Connes]{Alain Connes}
\address{A.~Connes: Coll\`ege de France \\
3, rue d'Ulm \\ Paris, F-75005 France
\\ I.H.E.S. and Vanderbilt
University} \email{alain\@@connes.org}

\date{}
\begin{document}
\maketitle \vspace{2cm}

\begin{center}
{\em  D\'edi\'e \`a Michel Dubois-Violette}
\end{center}

\medskip
\begin{verse}
On nous a tant assubjectis aux cordes \\
que nous n'avons plus  \\
de franches allures.
\\
\smallskip
Montaigne -- {\em Livre I, Chapitre 26}
\end{verse}

\begin{abstract} We introduce an invariant of Riemannian geometry which
measures the relative position of two von Neumann algebras in Hilbert space,
and which, when combined with the spectrum of the Dirac operator, gives a
complete invariant of Riemannian geometry. We show that the new
invariant plays the same role with respect to the spectral invariant as the
Cabibbo--Kobayashi--Maskawa
mixing matrix in the Standard Model  plays with respect to the list of
masses of the quarks.
\end{abstract}

\tableofcontents

\section{Introduction} It is well known since the original result of Milnor
\cite{Milnor}, that the spectral invariants \ie the spectrum of operators like
the Laplacian or the Dirac operator, are not complete invariants of Riemannian
geometry. The goal of this paper is to describe an additional invariant which
measures the relative position of two von Neumann algebras in Hilbert space,
and which, when combined with the spectrum of the Dirac operator, gives a
complete invariant of Riemannian geometry. We shall moreover show that the new
invariant plays the same role with respect to the spectral invariant as the
Cabibbo--Kobayashi--Maskawa (CKM)
matrix in the Standard Model \cite{mc2} plays with respect to the list of
masses of the quarks.

\smallskip

We shall first recall the role of the CKM matrix and explain, in \S \ref{sectckm}, its conceptual
meaning, whose mathematical side can in fact be traced back to the nineteenth century \cite{Sylvester},
\cite{Taber}, \cite{Autonne},
\cite{Browne}, \cite{EY}.   As we shall see, it encodes the relative position
of two maximal abelian von Neumann algebras $M$ and $N$ acting in three dimensional Hilbert space,
and the construction easily extends to the case of arbitrary finite dimension $n$.  At first the
invariant is  a unitary matrix $C_{xy}$ with line index $x\in \Sp(M)$
and column index $y\in \Sp(N)$. We then show how to eliminate, by considering the set of
lines of the matrix, the labeling
by the spectrum of $M$ and obtain an invariant $\Sp_N(M)$ of the relative position
of $M$ with respect to $N$ which is unaltered under automorphisms of $M$.

\smallskip

While our motivation comes from the example of the CKM matrix in the Standard Model another important source
of examples comes from Popa's theory of commuting squares \cite{Popa} which plays a central role in the
construction of subfactors. The special case of commuting squares means that the absolute values
$|C_{xy}|$ of all matrix elements $C_{xy}$ are equal (to $1/\sqrt N$) and this special case gives
rise to many interesting examples and questions (\cf \cite{Haagerup}, \cite{Nico}).

\smallskip

In \S \ref{contdisc} we
extend this CKM-invariant to the infinite dimensional case and
define  a complete invariant of the relative position of two commutative
von Neumann algebras, a discrete one and a
  continuous one, acting in the same (separable) Hilbert space.  By a result of
  von Neumann (\cf Theorem \ref{vNthm} below) there is, up to isomorphism, only one commutative von Neumann algebra $M$
  which is ``continuous" \ie has no minimal projection. Moreover, for a given multiplicity $m\in \N$ there
  is a unique (up to unitary equivalence) representation of $M$ as operators in Hilbert
  space with multiplicity $m$. The  position of $M$ relative to a discrete von Neumann algebra $N$
  (corresponding to diagonal operators for an orthonormal basis) is measured by the
  invariant $\Sp_N(M)$ which is a direct generalization of the CKM matrix. It is
  given by a measured section of a specific line bundle on a projective
  space of positive hermitian forms $\rho_{\lambda\kappa}$ of rank $m$ where $m$ is the multiplicity
  of the continuous von Neumann algebra $M$. The labels $\lambda,\kappa$ of the
  components of the hermitian  form are elements $\lambda,\kappa\in \Sp(D)$ in the
  spectrum of the discrete von Neumann algebra $N$. In the case of interest,
  the discrete von Neumann algebra is generated by an unbounded self-adjoint
  operator $D$ and the labels  can be viewed\footnote{modulo the issue of the multiplicity of eigenvalues.} as real
  numbers $\lambda,\kappa\in \Sp(D)\subset \R$.

  \smallskip

  The  whole information on a Riemannian
geometry can be encoded in a pair consisting of
\begin{itemize}
  \item The spectrum of the Dirac operator
  \item The relative position of two commutative von Neumann algebras, a discrete one and a
  continuous one.
\end{itemize}
The completeness of the invariant is closely related to
(and follows easily from) the embedding technique of \cite{bbg}.
The difficult part gives a characterization of the spectral triples thus obtained.
We shall then recall in
\S \ref{sectchar} the main result of \cite{CoRec} which gives a
characterization of the spectral triples corresponding to Riemannian
geometries.
We end with the computation (\S \ref{sectexamples}) of the invariant in a few examples, but this part is
still wanting for more convincing applications.

\section{The CKM matrix} \label{sectckm}

In the Standard Model giving the masses of the quarks does
not suffice to specify all parameters involved in their Yukawa couplings.
The missing parameters are the content of the CKM-mixing matrix which we now describe.
The weak isospin group $\SU(2)$, which is the gauge group
of the weak interactions, relates together the mass eigenstates of the up
quarks with those  of the down quarks and the corresponding basis (only given
up to phase) have a mismatch which is expressed by a {\em mixing} matrix
$C_{\lambda\kappa}$ whose indices $\lambda,\kappa$  label the three generations
of quarks. This matrix appears in the terms of the lagrangian such as:
\begin{center}
\begin{math}
\frac{ig}{2\sqrt{2}}W^{+}_{\mu}\left(
\bar{u}^{\lambda}_{j}\gamma^{\mu}(1+\gamma^{5})C_{\lambda\kappa}d^{\kappa}_{j}\right)+
\frac{ig}{2\sqrt{2}}W^{-}_{\mu}\left(\bar{d}^{\kappa}_{j}
C^{\dagger}_{\kappa\lambda}\gamma^{\mu}(1+\gamma^{5})u^{\lambda}_{j}\right)
\end{math}
\end{center}
It is responsible for the flavour-changing weak decays. The knowledge
of the CKM-matrix complements the list of quark masses
to specify the Yukawa couplings in the
Standard Model.
 The terminology
CKM-matrix refers to the names of N. Cabibbo who first treated the case of
two generations and of M. Kobayashi and T. Maskawa who treated the
case of three generations.

To be more specific for 2 generations, the matrix $C$ depends on just the
Cabibbo angle $\theta_c$, and is given by
\begin{equation}\label{CabAng}
C=\left[\begin{array}{cc} \cos\theta_c & \sin \theta_c \\
-\sin\theta_c & \cos \theta_c
\end{array}\right].
\end{equation}
For 3 generations the matrix $C$ has the more complicated form
\begin{equation}\label{CKMmatrix}
C= \left[ \begin{array}{ccc} C_{ud} & C_{us} & C_{ub} \\
C_{cd} & C_{cs} & C_{cb} \\
C_{td} & C_{ts} & C_{tb} \end{array} \right].
\end{equation}
The matrix $C$ depends upon the 3 angles $\theta_1,\theta_2,\theta_3$ and a complex
phase $\delta$. Let $c_i=\cos\,\theta_i$, $s_i=\sin\,\theta_i$, and $e_\delta=\exp(i\delta)$, then
\begin{equation}\label{CKMmatrixParam}
C= \left[ \begin{array}{ccc} c_1 & -s_1c_3 & -s_1s_3 \\
s_1c_2 & c_1c_2c_3 -s_2s_3 e_\delta & c_1c_2s_3 + s_2c_3 e_\delta \\
s_1s_2 & c_1s_2c_3+c_2s_3e_\delta & c_1s_2s_3-c_2c_3e_\delta
\end{array} \right].
\end{equation}

\smallskip
The mathematical treatment of the corresponding matrix problem
in arbitrary dimension goes back to the nineteenth century \cite{Sylvester},
\cite{Taber}, \cite{Autonne},
\cite{Browne}, \cite{EY}.
In this section we shall explain its conceptual meaning   in terms
which will then be extended to the infinite dimensional case.

\begin{figure}
\begin{center}
\includegraphics[scale=0.69]{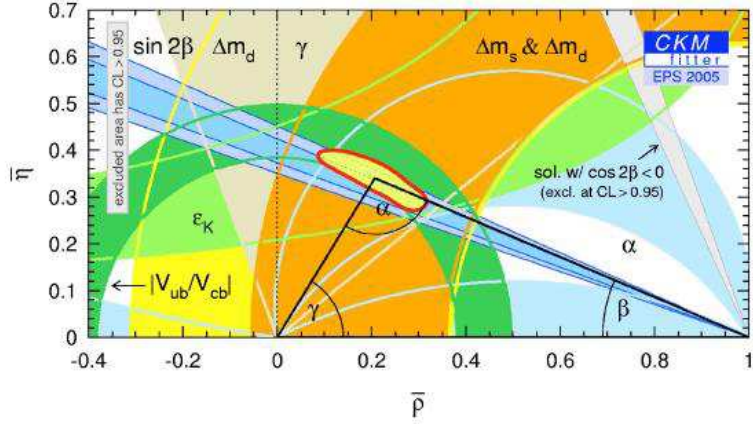}
\end{center}
\caption{Experimental information on the CKM matrix (noted $V_{xy}$
instead of $C_{xy}$). \label{pictckm} }
\end{figure}

\subsection{Pair of maximal abelian subalgebras with labeled idempotents}\hfill
\medskip

By construction the CKM matrix $C_{\lambda\kappa}$ conjugates two orthonormal
basis of the same three dimensional Hilbert space, but each of the basis
elements are only defined up to a phase. Thus $C_{\lambda\kappa}$ is a specific
representative of a double coset in the $4$-dimensional coset space
$$
\Delta_1\backslash \SU(3)/\Delta_1=\Delta\backslash {\rm U}(3)/\Delta
$$
where $\Delta_1$ (resp $\Delta$) is the group of unimodular diagonal matrices
in $\SU(3)$ (resp. of diagonal matrices in ${\rm U}(3)$). Up to the right
multiplication by the diagonal matrix $$ \left[
  \begin{array}{ccc}
    1 & 0 & 0 \\
    0 & -1 & 0 \\
    0& 0 & -1 \\
  \end{array}
\right] $$ and assuming $\theta_j\in ]0,\pi/2[$
the matrix $C_{\lambda\kappa}$ is the only element in the double
coset of $C$ in $\Delta\backslash {\rm U}(3)/\Delta$
whose first line and column are having positive entries. We shall now use
the same normalization condition in general.

\smallskip

Let $\cH$ be a Hilbert space of finite dimension $n$ and $M$, $N$ be two
(discrete) maximal abelian subalgebras of $\cL(\cH)$ with labeled minimal
idempotents $e_j\in M$, $f_j\in N$ for  $j\in \{1,\ldots, n\}$.

\begin{lem}\label{lemckm}  Assume that
$e_if_j\neq 0$ if $i$ or $j$ is $1$. Then there exists, up to an overall phase
factor, a unique pair of orthonormal basis $(\xi_j)$ and $(\eta_j)$ of $\cH$
such that
\begin{enumerate}
  \item[a)] $e_j\xi_j=\xi_j$ and $f_j\eta_j=\eta_j$ for all $j\in \{1,\ldots, n\}$.
  \item[b)]  $\langle \xi_j,\eta_i\rangle > 0$ if $i=1$ or $j=1$.
\end{enumerate}
There exists a unique unitary operator $U\in \cL(\cH)$ such that
$U\xi_j=\eta_j$ for all $j$. One has $Ue_j\,U^*=f_j$ for all $j\in \{1,\ldots,
n\}$.
\end{lem}

\proof Let us choose a unit vector $\xi_1$ with $e_1\xi_1=\xi_1$. Then there
exists, since $e_1f_1\neq 0$, a unique vector $\eta_1$ with $f_1\eta_1=\eta_1$
and such that $\langle \xi_1,\eta_1\rangle >0$. The conditions a) and b)
then uniquely fix the vectors $\xi_j$ and $\eta_j$ for $j>1$. Since the  pair
of orthonormal basis $\xi_j$ and $\eta_j$ is unique up to an overall phase the
operator $U$ is unique. The last assertion follows from $U\xi_j=\eta_j$.
\endproof

Let us check that the number of degrees of freedom is the right one. Assume
 given the positive scalars
\begin{equation}\label{scaldata}
    \alpha_j=\langle \xi_j,\eta_1\rangle\,,\ \ \beta_i=\langle
    \xi_1,\eta_i\rangle\,, \ \beta_1=\alpha_1\,, \ \ \sum \alpha_j^2=\sum
    \beta_j^2=1\,.
\end{equation}
This fixes the coordinates of $\eta_1$ in the given basis $\xi_j$. The only
constraints on $\eta_2$ are that $\Vert \eta_2\Vert =1$, $\eta_2\perp \eta_1$, $\langle
    \xi_1,\eta_2\rangle=\beta_2$. Thus the freedom of choice is a point in
    \begin{equation}\label{choice2}
      S_2=  \{\eta\in \cH\,|\,\Vert \eta\Vert =1\,, \ \eta\perp\eta_1\,, \ \langle
    \xi_1,\eta\rangle=\beta_2\}
    \end{equation}
and more generally one gets for the inductive choice of $\eta_k$,
\begin{equation}\label{choicek}
      S_k=  \{\eta\in \cH\,|\,\Vert \eta\Vert =1\,, \ \eta\perp\eta_j\,, \forall j<k\,, \ \langle
    \xi_1,\eta\rangle=\beta_k\}
    \end{equation}
    With $\dim \cH=n$ one gets that $S_k$ is the intersection of the
    unit sphere with a complex affine\footnote{By convention the inner product is antilinear
    in the first variable.} subspace of dimension $n-k$.
    Thus (if non-empty) its real dimension is $2(n-k)-1$.
    Thus the degrees of freedom come from the $2n-1$ scalars
    \eqref{scaldata} which fulfill $2$ conditions and hence give
    real dimension $2n-3$ and then the sum of the $2(n-k)-1$ for $k$
    between $2$ and $n-1$. This gives $(n-1)^2$ real parameters as
    expected from the dimensions $n^2-1$ of ${\rm SU}(n)$ and $n-1$ of the group
    of diagonal unimodular unitaries.

\subsection{Generic representations of the free product $M\star N$}\hfill
\medskip\label{freeprod}

  Let $M$ and $N$ be  commutative von Neumann algebras of dimension
    $n<\infty$ and $e_M\in M$, $e_N\in N$ minimal idempotents, or equivalently elements of
    the spectrum. We let  $M\star N$ be the $C^*$-algebra free product of $M$
    and $N$. Thus giving a Hilbert space representation of  $M\star N$ is the
    same as giving a pair of representations of $M$ and $N$ in the same Hilbert
    space.
\begin{defn}
    We say that a
    representation  $\pi$ of the free product $M\star N$ in a Hilbert space
    $\cH$ of dimension $n$ is {\em generic} when
the states $y\mapsto \Tr(\pi(e_M \,y))$ on $N$ and $x\mapsto \Tr(\pi(x\,
      e_N))$ on $M$ are faithful.
    \end{defn}
    This implies that the restriction of $\pi$ to $M$ and $N$ are
    isomorphisms with maximal abelian subalgebras of $\cL(\cH)$.

\begin{prop}\label{propckm} The generic representations
of the free product $M\star N$ in a Hilbert space
    $\cH$ of dimension $n$ are  classified up to unitary
equivalence by matrices $C_{xy}$ with $x\in \Sp(M)$, $y\in\Sp(N)$
 such that
\begin{itemize}
  \item $C$ is a unitary matrix
  \item The line and column of $C$ corresponding to $e_M$ and $e_N$ are
  strictly positive.
\end{itemize}
\end{prop}

\proof Let us first show how to associate a matrix $C(\pi)$ to a generic
representation $\pi$ of the free product $M\star N$. We choose two unit vectors
$\xi,\eta \in \cH$ such that:
\begin{equation}\label{norma}
    \pi(e_M)\xi=\xi\,, \ \ \pi(e_N)\eta=\eta\,, \ \ \langle
    \xi,\eta\rangle >0\,.
\end{equation}
For each $x\in \Sp(M)$ there is a unique corresponding minimal projection
$e_x\in M$ and a unique unit vector $\xi_x\in \cH$ such that $e_x\xi_x=\xi_x$
while $\langle
    \xi_x,\eta\rangle >0$. A similar statement holds for $N$, with
    $f_y\eta_y=\eta_y$ while $\langle
    \xi,\eta_y\rangle >0$. We then define the matrix $C$ as follows
\begin{equation}\label{defnC}
    C_{xy}=\langle
    \xi_x,\eta_y\rangle \qqq x\in \Sp(M), y\in \Sp(N)\,.
\end{equation}
The choice of the pair $(\xi,\eta)$ is only unique up to an overall phase \ie up
to the modification:
$(\xi,\eta)\to(\lambda \xi,\lambda\eta)$ with $|\lambda|=1$. All the vectors
$\xi_x$ and $\eta_y$ get multiplied by $\lambda$ and this does not affect
\eqref{defnC}. This shows that the matrix $C$ is an invariant of the
representation $\pi$. One has
$$
\sum_x \overline C_{xy_1} C_{xy_2=}=\sum_x \langle
    \eta_{y_1},\xi_x\rangle \langle
    \xi_x,\eta_{y_2}\rangle=\langle
    \eta_{y_1},\eta_{y_2}\rangle=\delta_{y_1,y_2}
$$
and similarly
$$
\sum_y C_{x_1y} \overline C_{x_2y} =\sum_y  \langle
    \xi_{x_1},\eta_y\rangle\langle
    \eta_y,\xi_{x_2}\rangle=\langle
    \xi_{x_1},\xi_{x_2}\rangle=\delta_{x_1,x_2}\,.
$$
This shows that the matrix $C$ is unitary.  The second property of lines and
columns of $C$ is true by construction. Let us first show that $C(\pi)$ is a
complete invariant. For this we define a ``model" $\pi_C$ for each $C$ and show
that the given representation $\pi$ is unitarily equivalent to $\pi_C$ for
$C=C(\pi)$. This will also show that all invariants $C$ are obtained.

To construct $\pi_C$ we take $\ell^2(\Sp(N))$ with the diagonal action of $N$,
and we define vectors $\zeta_x$ indexed by $x\in \Sp(M)$ by
\begin{equation}\label{modcons}
\langle
    \zeta_x,\epsilon_y\rangle=C_{xy}\qqq y\in \Sp(N)
\end{equation}
where $\epsilon_y$ is the canonical basis of $\ell^2(\Sp(N))$. The vectors $\zeta_x$ form an
orthonormal basis of $\ell^2(\Sp(N))$ and one can thus define the action of $M$
on $\ell^2(\Sp(N))$ as the diagonal action in this basis. Combining this
with the canonical representation of $N$ in $\ell^2(\Sp(N))$ we thus obtain a
representation $\pi_C$ of the free product $M\star N$ in the Hilbert space
    $\ell^2(\Sp(N))$. One checks using \eqref{modcons} that
     the corresponding $C(\pi_C)$ is $C$.

Starting from a given representation $\pi$ in $\cH$ we use Lemma \ref{lemckm}
and get the corresponding orthonormal basis $\xi_x$ and $\eta_y$ as above. We
let $V$ be the unitary map from $\cH$ to $\ell^2(\Sp(N))$ with
$V\eta_y=\epsilon_y$. One then has, using \eqref{defnC} and \eqref{modcons}
that $V\xi_x=\zeta_x$ which gives the unitary equivalence $\pi\sim
\pi_{C(\pi)}$.\endproof

\smallskip

\begin{rem} \label{rempara}{\rm  Note that the matrix $C$ is relating two different spaces $\ell^2(\Sp(M))$
and $\ell^2(\Sp(N))$. Indeed there is no a priori identification of $\Sp(M)$
with $\Sp(N)$. Note also that the square of the  absolute value $|C_{xy}|^2$ is
given by the simple expression $|C_{xy}|^2=\Tr(e_xf_y)$ in terms of the minimal
projections of $M$ and $N$. In general the knowledge of the absolute values $|C_{xy}|$ does not
suffice to recover $C$ and for instance the case $|C_{xy}|^2=1/n$ for all $x,y$ corresponds to
the mutually commuting case of \cite{Popa}, Definition 2.2.}
\end{rem}

\subsection{The relative spectrum $\Sp_N(M)$}\hfill
\medskip

In the construction of \S \ref{freeprod}, the entries of the  matrix $C$  are
labeled by the spectra of $M$ and $N$. We shall now show how to eliminate the
labeling coming from the spectrum of $M$. The resulting invariant will give the
relative position of $M$ viewed as a subalgebra of $\cL(\cH)$ relative to $N$
viewed as a von Neumann algebra with known spectrum. In order to eliminate the
parameter $x\in \Sp(M)$ we take the range of the function from $\Sp(M)$ to
$N$ which assigns to $x\in \Sp(M)$ the corresponding line $C_{xy}, y\in \Sp(N)$
of the matrix $C$. Since $N$ is canonically the algebra of complex valued functions
on $\Sp(N)$ one can view a line  $(C_{xy})_{ y\in \Sp(N)}$ as an element of $N$
which we denote by $C_{x,\,\bullet}$.
We define
\begin{equation}\label{rangeC}
    G(\pi)=\{  C_{x,\,\bullet}\in N\,|\, x\in \Sp(M)\}\subset
    N\,.
\end{equation}
Since $C$ is a unitary matrix  its lines belong to the
unit sphere  in $N$ defined
by
\begin{equation}\label{sphere}
S^N_\C=\{z\in N\,|\,\sum_y |z_y|^2=1\}
\end{equation}
Thus  $\Sigma=G(\pi)$ is a subset of the unit sphere. One
recovers the matrix $C$ from the subset just by taking the coordinates of its
$n$ elements. This is not yet invariantly defined since we needed the base
points $e_M$ and $e_N$ as well as the generic condition. In order to eliminate
this choice of base points, we consider the action, by multiplication, of the
unitary group $\cU(N)$ on the projective space $\P_N=S^N_\C/{\rm U}(1)$. We let
$p:S^N_\C\to \P_N$ be the projection.
\begin{defn} \label{defnrelspec} We define the {\em relative spectrum}
of $M$ relative to $N$ as
\begin{equation}\label{relspec}
    \Sp_N(M)=p(G(\pi))\subset \P_N\,,
\end{equation}
viewed as a subset defined up to the gauge action of the unitary group
$\cU(N)$.
\end{defn}
The gauge ambiguity shows up in the choice of an isomorphism of the representation of
$N$ in $\cH$ with the canonical representation of $N$ in $\ell^2(\Sp(N))$. Once this
choice is done, the minimal projections of $M$ are just elements of $\P_N$.
One can partially  fix the gauge in the generic case by requiring that one of the
elements of $\Sp_N(M)$ has all coordinates $>0$.

\smallskip We shall now give an equivalent description of $\P_N$ which will allow one to treat
the case where the action of $M$ has multiplicity $>1$.
For each $k$, let $\her_k(N)$
be the space of rank $k$ positive hermitian matrices $\rho_{\lambda\mu}$ with labels
$\lambda,\mu\in \Sp(N)$.  One way to encode a point in $y\in\P_N$ is by the
corresponding rank one  matrix $\rho\in \her_1(N)$,
\begin{equation}\label{rankone}
    \rho_{\lambda\mu}=\bar z_\lambda z_\mu\, , \ \ z\in S^N_\C\,, \  p(z)=y
\end{equation}
We can then view $S=\Sp_N(M)$ as a subset with $n$ elements in $\her_1(N)$ such, for the
matrix product, the following conditions hold:
\begin{itemize}
  \item $\rho^2=\rho$ for all $\rho\in S$
  \item $\rho \rho'=0$ for all $\rho\neq \rho'\in S$
  \item $\sum_S \rho_{\lambda\mu}=\delta_{\lambda\mu}$
\end{itemize}
This  is just encoding the partition of unity in the rank one minimal
projections of $M$. In these coordinates the gauge action of the unitary group
$\cU(N)$ becomes the adjoint action:
\begin{equation}\label{adjact}
   ( {\rm Ad}(u)\rho)_{\lambda\mu}=  u_\lambda \rho_{\lambda\mu}\bar u_\mu
\end{equation}
It might seem that replacing the subset $\Sp_N(M)=p(G(\pi))\subset \P_N$ by the
subset of $\her_1(N)$ is a useless complication but it allows one to extend the definition and properties of
$\Sp_N(M)$ to
the case when the abelian von Neumann algebra $M$ is no longer maximal abelian.
We fix its multiplicity\footnote{It can be easily extended to the general case
of non-constant multiplicity.} to be constant equal to $m$. We then get in the
Hilbert space $\ell^2(\Sp(N))$ a partition of unity in the rank $m$ minimal
projections of $M$. They form a subset with $n/m$ elements  $S\subset
\her_m(N)$ and fulfill exactly the same rules as above. We thus get:

\begin{prop} \label{lempartunit} Let $N$ be a maximal abelian von Neumann subalgebra of $\cL(\cH)$
with $\cH$ of dimension $n$. The relative position of abelian von Neumann
algebras of constant multiplicity $m$ in $\cL(\cH)$ is classified by subsets
$S=\Sp_N(M)\subset \her_m(N)$ with $n/m$ elements such that (for the matrix product)
\begin{itemize}
  \item $\rho^2=\rho$ for all $\rho\in S$
\item $\rho \rho'=0$ for all $\rho\neq \rho'\in S$
  \item $\sum_S \rho_{\lambda\mu}=\delta_{\lambda\mu}$
\end{itemize}
The subset $S$ modulo the adjoint action of the unitary group $\cU(N)$ is a
complete invariant.
\end{prop}

\proof This is a simple reformulation of the equivalence between giving $M$ and
giving the partition of unity by its minimal projections in the Hilbert space
$\ell^2(\Sp(N))$ which is isomorphic to $\cH$ with an isomorphism which is
unique modulo the unitary group $\cU(N)$.\endproof

\subsection{The CKM matrix and Fourier transform}\hfill \medskip

Let us compute the above invariant $\Sp_N(M)$ in a simple example. We
 consider a finite abelian group $G$ and the pair of maximal abelian subalgebras of $\ell^2(G)$
  given by the algebra $N=\ell^\infty(G)$
of multiplication operators and the algebra $M$ of convolution operators. We let
$G=\Z/3\Z$ as a concrete example. We take the base points given for the algebra $N$ by the delta function at
$1\in G$ and for $M$ by the idempotent $e_M(g)=1$ of the
convolution\footnote{We use the normalized Haar measure of total mass $1$.}
algebra $C^*(G)$. The corresponding
CKM matrix is of the form
$$
C=
\frac{1}{\sqrt 3}\left[
  \begin{array}{ccc}
    1 & 1 & 1 \\
    1 & e^{\frac{2\pi i}{3}} & e^{-\frac{2\pi i}{3}} \\
    1 & e^{-\frac{2\pi i}{3}} & e^{\frac{2\pi i}{3}}\\
  \end{array}
\right]
$$
which gives with the notations of \eqref{CKMmatrixParam},
$$
c_1=\frac{1}{\sqrt 3},\  s_1=\sqrt\frac{ 2}{ 3},
\ c_2=s_2=\frac{1}{\sqrt 2}, \ c_3=s_3=-\frac{1}{\sqrt 2}\,, \ \ e_\delta=i\,,
$$
The relative spectrum $\Sp_N(M)$ in the sense of Definition \ref{defnrelspec}
is given by the  three points in $\P_2(\C)$ given by the lines of the above matrix. These three
points are pairwise orthogonal and the gauge fixing corresponds to one of them (namely $(1,1,1)$)
having all its coordinates positive. Note that $\Sp_N(M)$ is invariant under the complex
conjugation but the CKM matrix itself is not, since $e_\delta=i$. This corresponds to
the nuance with the more refined invariant of $M$ where the labeling of the spectrum is specified.

\begin{rem} \label{comsquare}
{\rm The above example falls in the special class of {\em commuting squares} (\cf \cite{Popa}).
More precisely (\cf \cite{Popa}, Definition 2.2) a pair of
von Neumann subalgebras $(M,N)$ of a finite von Neumann algebra $P$ with normalized trace $\tau$
is called mutually orthogonal when
\begin{equation}\label{compair}
    \tau(b_1b_2)=\tau(b_1)\tau(b_2)\qqq\, b_1\in M \,, \ b_2\in N\,.
\end{equation}
This condition is equivalent in the above context of a pair of maximal abelian
von Neumann subalgebras $(M,N)$ of $M_n(\C)$ to $|C_{xy}|=1/\sqrt n$ for all $x,y$.

} \end{rem}
\smallskip

\section{Relative position of the continuum and the discrete} \label{contdisc}

Let us first recall the following result of von Neumann (\cite{vNeumann}, \cite{vNeumann1})  which shows that
there is a unique way to represent the continuum with constant multiplicity $m$.

\begin{thm}\label{vNthm}
Let $\cH$ be an infinite dimensional separable\footnote{\ie with countable
orthonormal basis.} Hilbert space and $m $ an integer. There exists up to
unitary equivalence only one commutative von Neumann subalgebra $M\subset
\cL(\cH)$ such that,
\begin{enumerate}
  \item $M$ contains no minimal projection,
  \item The commutant of $M$ is isomorphic to $M_m(M)$.
\end{enumerate}
\end{thm}

\proof We briefly recall the proof
of Theorem \ref{vNthm}.
The meaning of (1) is that $M$ represents the continuum, while the meaning of
(2) is that the multiplicity of $M$ in $\cH$ is equal to $m$.
The separability of $\cH$ shows that $M$ is generated\footnote{As a von Neumann algebra.}
 by a countable collection $T_k$
of commuting self-adjoint operators and hence by a single self-adjoint
operator $T=f_\infty(T_1,\ldots,T_k,\ldots)$  where $f_\infty$ is a Borel
injection of an infinite product of intervals in $\R$ into the real line $\R$ so that
$T_k$ is a Borel function $f_k(T)$ of $T$ for each $k$. The multiplicity hypothesis shows that
the representation of $M$ in $\cH$ is the direct sum of $m$ copies of the representation
in $e\cH=\cH_1$, $e$ a minimal projection in $M_m(\C)\subset M_m(M)$. Thus one can assume
that $m=1$. In that case $M$ has a cyclic and separating unit vector $\xi$ and the spectral Theorem
shows that the operator $T$ is unitarily equivalent to the operator of multiplication
by $x$ in $\cH_1=L^2(K,d\nu)$ where $K\subset \R$ is the compact spectrum of $T$ and
$d\nu$ the spectral measure,
$$
\int f(x)d\nu(x)=\langle \xi,f(T)\xi\rangle\,.
$$
Since $M$ contains no minimal projection, the function $\nu(u)=\int_{-\infty}^ud\nu(x)$
is continuous. One has, for $u<v\in \R$,
$$
\nu(u)\leq \nu(v)\,, \ \ \nu(u)=\nu(v)\Leftrightarrow [u,v]\cap K=\emptyset
$$
Thus $\nu$ is injective \alm for $d\nu$, the operator $\nu(T)$ generates $M$ and is unitarily
equivalent to the operator of multiplication by $x$ in $L^2([0,1],dx)$. It follows that the
pair $(M,\cH)$ is unitarily equivalent to the pair
\begin{equation}\label{modelvN}
(L^\infty(X,\mu), L^2(X,\mu,S))\,, \ \ X=[0,1]\,,\  \mu=dx\,, \ S=\C^m\,,
\end{equation}
where the trivial bundle with fibers $S_x=\C^m$ is viewed as  a measurable hermitian vector
bundle $S$ of dimension $m$ over $X$, and the action of $M=L^\infty(X,\mu)$ is given
by multiplication in the Hilbert $L^2(X,\mu,S)$ of $L^2$-sections of $S$.
\endproof

It is useful in general to keep the flexibility of describing the above model
of the pair $(M, \cH)$ using \eqref{modelvN} with $(X,\mu)$ a standard probability space and
 a measurable hermitian vector
bundle $S$ of dimension $m$ over $X$. This is unique up to
equivalence (\cite{vNeumann} \S 9 Definition 3).

\smallskip
Our goal in this section is to extend the result of \S \ref{sectckm} and get a
complete invariant of the relative position in $\cH$ of the pair $(M,N)$ where
$M$ is as in Theorem \ref{vNthm} and $N$ is a discrete maximal abelian von Neumann subalgebra $N\subset \cL(\cH)$.
By discrete we mean that $N$ is generated by its minimal projections. In that
case it is isomorphic to $\ell^\infty(\N)$ acting in $\ell^2(\N)$ by
multiplication. We shall later adapt the result to the case when $N$ has
multiplicity and this will not introduce any new major difficulty.

\subsection{The Frechet manifolds $\her_m(\Lambda)$, $\per_m(\Lambda)$ and $\gra_m(\Lambda)$} \hfill \medskip

We let $\Lambda$ be a countable set, which will label the  set
$\Sp(N)$ of minimal projections of the discrete von Neumann algebra $N$.
Let $\her_m(\Lambda)$ be the space of positive hermitian forms of rank $m$ of the
form $\rho_{\lambda\mu}$ where the indices $\lambda,\mu$ belong to $\Lambda$. Saying
that $\rho_{\lambda\mu}$ is of rank $m$ means that the separated completion of the
space $c_c(\Lambda)$ of sequences with finite support on $\Lambda$ for the inner
product,
\begin{equation}\label{innprod}
    \langle a,b\rangle=\sum \rho_{\lambda\mu} \bar a_\lambda b_\mu \qqq a, b \in c_c(\Lambda)
\end{equation}
is a finite dimensional Hilbert space $\Ss(\rho)$ of dimension $m$. Note that the
completion is unnecessary since a dense subspace of a finite dimensional
Hilbert space is equal to the Hilbert space. Thus $\Ss(\rho)$ is the quotient
\begin{equation}\label{radical}
\Ss(\rho)=c_c(\Lambda)/J_\rho\,, \ \ J_\rho=\{a\in c_c(\Lambda)\,|\,\sum \rho_{\lambda\mu}
\bar a_\lambda a_\mu=0\}\,.
\end{equation}
By construction the radical $J_\rho$ of $\rho$ is a codimension $m$ subspace of $c_c(\Lambda)$.
It can be equivalently described by orthogonality with the lines of $\rho$ \ie
\begin{equation}\label{radicalbis}
 J_\rho=\{a\in c_c(\Lambda)\,|\,\sum \rho_{\lambda\mu}
  a_\mu =0 \qqq \lambda\in \Lambda\}\,.
\end{equation}
We let $\gra_m(\Lambda)$ be the Grassmanian of all codimension $m$ subspaces $J$ of $c_c(\Lambda)$.
It is described equivalently as the Grassmanian of all $m$-dimensional subspaces $H$ of
the Frechet space $\C^\Lambda$
using the canonical duality between $c_c(\Lambda)$ and $\C^\Lambda$ and the map $J\mapsto H=J^\perp$.
 We can use Grassmann
coordinates
\begin{equation}\label{grasscoor}
    H\mapsto \pi(H)=\xi_1\wedge \cdots \wedge \xi_m \qqq (\xi_j)\,\ {\rm basis\; of } \ H
\end{equation}
and view $\gra_m(\Lambda)$
as a subset of the projective space $\P(\wedge^m \C^\Lambda)$ over the linear space $\wedge^m \C^\Lambda$. There is a
natural duality between $\wedge^m \C^\Lambda$ and $\wedge^m c_c(\Lambda)$ and we endow $\wedge^m \C^\Lambda$
 with the weak topology coming from this duality. One
can identify $\wedge^m c_c(\Lambda)$ with $ c_c(\wedge^m\Lambda)$ where we let $\wedge^m\Lambda$
denote the set of oriented subsets $F\subset \Lambda$ with $m$ elements.
The range of $\pi$ is characterized by the Pl\"{u}cker relations: $i_VP\wedge P=0$ for the contraction
of $P=\pi(H)$ with any $V\in \wedge^{m-1}c_c(\Lambda)$.
A local coordinate chart in $\gra_m(\Lambda)$ around $H$ is obtained by choosing a
closed supplement $H'$ of $H$ in $\C^\Lambda$, \eg with $H=J^\perp$ by taking an $m$-dimensional
 supplement $J'$ of $J$
in $c_c(\Lambda)$ and letting $H'$ be the orthogonal of $J'$.
The domain of the local chart is the set of $m$-dimensional subspaces of $\C^\Lambda$
 which intersect $H'$ trivially.
The local chart describes these subspaces as the graphs of arbitrary linear maps
$T\,:\, H\to H'$. Let $p(H,H')$ be the projection on $H$ parallel to $H'$. The
 change of charts from $(H_1,H'_1)$ to  $(H_2,H'_2)$ is given by
 \begin{equation}\label{changechart}
 T_2=(1+T_1)R(T_1)-1
 \end{equation}
  where $R(T_1)$ is the
 inverse of the map
 $\xi\in H_1\mapsto p(H_2,H'_2)(\xi+T_1\xi)\in H_2$. Indeed for $\eta\in H_2$ one has $\eta+T_2\eta=(1+T_1)\xi$
 for $\xi=R(T_1)\eta \in H_1$. Moreover $T_2\eta=(1-p(H_2,H'_2))(1+T_1)\xi\in H'_2$.
 By construction \eqref{changechart} only involves inverses of linear maps acting in finite dimensional
spaces. Thus $\gra_m(\Lambda)$ is a Frechet manifold.

\begin{prop}\label{structgrass} 1) The  quotient $\Ss(J)=c_c(\Lambda)/J$ yields a vector bundle $\Ss$ of
dimension $m$ over $\gra_m(\Lambda)$.

2) The map $\rho\in \her_m(\Lambda)\mapsto j(\rho)=J_\rho\in \gra_m(\Lambda)$ is
a locally trivial fibration with fiber the $m^2$-dimensional real open cone of non-degenerate
positive hermitian forms on the corresponding fiber of the vector bundle  $\Ss$.
\end{prop}

\proof 1) On the domain of a local chart associated to the pair $(H,H')$ as above,
the projection on $H$ parallel to $H'$ gives a local trivialization of the bundle $\Ss$. The
 change of charts from $(H_1,H'_1)$ to  $(H_2,H'_2)$ is given by $T_1\mapsto R(T_1)$ which is
 a smooth map to invertible linear maps from $H_2$ to $H_1$.

2) Let $J\subset c_c(\Lambda)$ be a subspace of codimension $m$ and $\Ss(J)=c_c(\Lambda)/J$.
Giving $\rho\in \her_m(\Lambda)$ such that $j(\rho)=J$ is equivalent to choosing a
 non-degenerate
positive hermitian form $h$ on $\Ss(J)$. One lets
$\rho_{\lambda\kappa}=\langle \delta_\lambda,\delta_\kappa\rangle_h$ for all $\lambda,\kappa\in \Lambda$.

\endproof

 We
denote by $\per_m(\Lambda)$ the quotient of $\her_m(\Lambda)$ by the scaling action of
$\R_+^*$,
\begin{equation}\label{projperspace}
    \per_m(\Lambda)=\her_m(\Lambda)/\R_+^*\,, \ \ \R_+^*\to \her_m(\Lambda)\stackrel{q}{\to} \per_m(\Lambda)\,.
\end{equation}

\begin{prop}\label{structproj} 1) The scaling action of $\R_+^*$ turns
$\her_m(\Lambda)$ into a principal $\R_+^*$-bundle $L$ over $\per_m(\Lambda)$.

3) A section $\sigma$ of the restriction of $L$ to $Y\subset \per_m(\Lambda)$ gives a
hermitian metric on the restriction of $\Ss$ to $Y$ by
\begin{equation}\label{metricherm}
    \langle \xi,\eta\rangle_y=\sum \rho_{\mu\nu}(\sigma(y))\bar{\xi}_\mu\eta_\nu\qqq \xi,\eta\in c_c(\Lambda)
\end{equation}

\end{prop}

\proof The proof is straightforward using Proposition \ref{structgrass}.
\endproof

\smallskip

\subsection{Injectivity of the map to hermitian forms} \hfill \medskip

As explained above, there is, up to measurable isomorphism, a unique model of the inclusion
$M\subset \cL(\cH)$. It is given by a standard probability space $(X,\mu)$ and
a measurable hermitian vector bundle $S$ of dimension $m$ over $X$, and the
action of $M=L^\infty(X,\mu)$ by multiplication in the Hilbert $L^2(X,\mu,S)$
of $L^2$-sections of $S$.

\begin{lem} \label{injecphi} Let $\cH$ be the Hilbert space $L^2(X,\mu,S)$ and
 $(\xi_n)_{n\in \N} $  an orthonormal basis.
Then the map $\gamma$ which to $x\in X$ associates the matrix
\begin{equation}\label{matrixgammamap}
\gamma(x)=\gamma_{n,m}(x)=\langle
\xi_n(x),\xi_m(x)\rangle
\end{equation}
  is injective outside a null-set even after moding out
by the scaling action of $\R_+^*$.
\end{lem}

\proof Let $\xi=\sum a_n\xi_n\in \cH$. The sequence $(a_n)$ belongs to
$\ell^2(\N)$. Let $m_k\in \N$ be such that
$$
\sum_{m_k}^\infty |a_n|^2\leq 2^{-2k}\,.
$$
One then has, with $\zeta_k=\sum_1^{m_k}a_n\xi_n$, that $\Vert \xi
-\zeta_k\Vert\leq 2^{-k}$ and thus except on a subset $E_k\subset X$ of measure
$\mu(E_k)\leq 2^{-k}$ one has $\Vert\xi(x) -\zeta_k(x)\Vert^2\leq 2^{-k}$. Thus
by the Borel-Cantelli Lemma, for almost all $x\in X$ one has $x\notin E_k$ except
for finitely many values of $k$ and hence $\zeta_k(x)\to
\xi(x)$. This shows that for almost all $x\in X$
$$
\Vert \xi(x)\Vert^2=\lim_{k\to\infty}\sum_{n=1}^{m_k}\sum_{n'=1}^{m_k}\bar a_n
a_{n'}\gamma_{n,n'}(x)\,.
$$
In particular it shows that, for any two given vectors $\xi,\eta\in \cH$,
 the knowledge of $\gamma_{n,m}(x)$ suffices to
determine almost everywhere the inner product $\langle \xi(x),\eta(x)\rangle$.
 This shows that the problem of
injectivity outside a null-set of the map $x\mapsto \gamma(x)$ is in fact
independent of the choice of the orthonormal basis $(\xi_n)_{n\in \N} $. More precisely
let $(\xi_n)_{n\in \N} $ be an orthonormal basis such that $q\circ\gamma$ is injective in
the complement of the null set $X_0\subset X$. Let $(\xi'_n)_{n\in \N} $ be another orthonormal basis.
Then by the above argument one can express the matrix elements $\gamma_{n,m}(x)$ as
a pointwise limit of linear functions $$
\gamma_{n,m}(x)=\lim_r L^{(r)}_{n,m}(\gamma'_{k,\ell}(x))$$ of the $\gamma'_{k,\ell}(x)$
except on a null set $X_1\subset X$. It follows that the map $q\circ\gamma'$ is injective
except on the null set $X_0\cup X_1$ since the proportionality
$$
\gamma'_{k,\ell}(x)=\lambda \gamma'_{k,\ell}(y)\qqq k,\ell
$$
implies  $\gamma_{n,m}(x)=\lambda\gamma_{n,m}(y)$ for all $n,m$.
  Up to
isomorphism we can   assume that the triple $(X,\mu,S)$ is given by the
circle $X=S^1$ with the measure $d\theta$ and that the bundle $S$ is the
trivial bundle. We first take the case where $S$ is of dimension one. Then we
have $L^2(X,\mu,S)=L^2(S^1,d\theta)$ and we can choose the basis given by
$\xi_n(\theta)=e^{in\theta}$ labeled by $n\in \Z$. The map $\gamma$ is then
$$
\gamma_{n,m}(\theta)=e^{i(m-n)\theta}\qqq m,n\in \Z\,.
$$
A relation of proportionality $\gamma(\theta)=\lambda\gamma(\theta')$ means
$$
e^{i(m-n)\theta}=\lambda e^{i(m-n)\theta'}\qqq m,n\in \Z\,.
$$
Taking $m=n$ gives $\lambda=1$ and then for $m=1$, $n=0$ one gets that $\theta
=\theta'$ (modulo $2\pi$). With $S$ trivial of dimension $\ell$ we take a basis of
the form $\xi_{n,k}(\theta)=e^{in\theta}\epsilon_k$ where the $\epsilon_k$ form
an orthonormal basis of $S$. One then has
$$
\gamma_{(n,k),(m,\ell)}(\theta)=e^{i(m-n)\theta}\delta_{k,\ell}
$$
and the map $\gamma$ is injective  even after moding out by a scaling factor,
since this already holds for the components $\gamma_{(n,1),(m,1)}$.\endproof

\subsection{The relative spectrum $\Sp_N(M)$}\hfill
\medskip

A discrete maximal abelian von Neumann subalgebra $N$ of $\cL(\cH)$ is isomorphic
to the algebra $\ell^\infty(\Lambda)$ acting by multiplication
 in $\ell^2(\Lambda)$ where $\Lambda=\Sp(N)$ is the
countable set of minimal projections of $N$.

Let  $F:L^2(X,\mu,S)\to \ell^2(\Lambda)$ be a unitary isomorphism.  Let
$\epsilon_\lambda$, $\lambda\in \Lambda$, be the canonical basis of
$\ell^2(\Lambda)$. The vectors $\eta_\lambda=F^*\epsilon_\lambda$ form an
orthonormal basis of $L^2(X,\mu,S)$. Each of them is a measurable $L^2$-section
$\eta_\lambda(x)$ of $S$ on $X$ and is defined almost everywhere modulo $\mu$.
Thus the following map is well defined almost everywhere modulo $\mu$,
\begin{equation}\label{mapvarphi}
\varphi(x)\in \her_m(\Lambda)\,, \ \ \varphi(x)_{\lambda\kappa}=\langle
\eta_\lambda(x),\eta_\kappa(x)\rangle
\end{equation}

Lemma \ref{injecphi} shows that this map is injective outside a null-set even
after composition with the quotient map $q$, \ie passing to $\per_m(\Lambda)$.

\begin{defn}\label{defngen} A measured section  of the bundle
 $L$ over $\per_m(\Lambda)$ is an equivalence class
of pairs $(\nu,\xi)$ of a positive finite measure $\nu$ on $\per_m(\Lambda)$ and a
section $\xi$ of $L$ defined almost everywhere for $\nu$, modulo the
equivalence relation
\begin{equation}\label{equiv}
    (\nu,\xi)\sim (h\nu,h^{-1}\xi)\qqq h:\per_m(\Lambda)\to \R_+^*\,.
\end{equation}
\end{defn}

We work as above in the measurable category, so that the function $h$ in
\eqref{equiv} is a measurable function.

\begin{lem}\label{lemind} 1) The pair $(\nu,\xi)$ where
\begin{equation}\label{gencons}
    \nu=(q\circ\varphi)_*(\mu)\,,\ \ \xi(q\circ\varphi(x))=\varphi(x)
\end{equation}
is a measured section  of $L$.

2) The measured section  of $L$ given by \eqref{gencons} only depends upon the von Neumann
subalgebra $M=FL^\infty(X,\mu)F^*\subset \cL(\ell^2(\Lambda)) $.
\end{lem}

\proof 1) By Lemma \ref{injecphi}, the map $q\circ\varphi$ is injective and
thus the equation $\xi(q\circ\varphi(x))=\varphi(x)$ determines a section $\xi$
of $L$ on its range $Y=q\circ\varphi(X)$. Thus the pair $(\nu,\xi)$ is a measured section
of $L$.

2) Let $X',\mu', S', F'$ be such that, as for $F$, the unitary $F'$ is an
isomorphism $F':L^2(X',\mu',S')\to \ell^2(\Lambda)$. Assume that, as subsets of
$\cL(\ell^2(\Lambda))$,
$$
FL^\infty(X,\mu)F^*=F'L^\infty(X',\mu')F'^*\,.
$$
Then the unitary $F^*F'$ conjugates the von Neumann algebra
$L^\infty(X',\mu')$ with $L^\infty(X,\mu)$. Thus (\cite{dix} A-85, \cite{Pedersen} Theorem 4.11.9)
 there exists a measurable isomorphism (outside null sets) $\psi:X\to X'$, a
measurable function $h:X\to \R_+^*$ and an isomorphism $V$ of measurable
hermitian bundles $\psi^*(S')\sim S$ such that
\begin{equation}\label{measuriso}
\psi_*(h\mu)=\mu'\,, \ \ F^*F'(\eta)(x)=h(x)^{1/2}V(x)\eta(\psi(x))\,.
\end{equation}
The isometric property of $U=F^*F'$ relates to the unitarity of $V$ by
$$
\int_X\Vert F^*F'(\eta)(x)\Vert^2d\mu(x)=\int_X\Vert
\eta(\psi(x))\Vert^2h(x)d\mu(x)=\int_{X'}\Vert \eta(x')\Vert^2d\mu'(x')\,.
$$
One has, with $U=F^*F'$,
\begin{equation}\label{isovonn}
    UfU^*=f\circ\psi \qqq f\in L^\infty(X',\mu')\,.
\end{equation}
Let, as above, $\eta_\lambda=F^*\epsilon_\lambda$ and
$\eta'_\lambda=F'^*\epsilon_\lambda$. One has $\eta_\lambda=U\eta'_\lambda$ and
thus
\begin{equation}\label{pointwisev}
\eta_\lambda(x)=h(x)^{1/2}V(x)\eta'_\lambda(\psi(x))
\end{equation}
which since $V(x)$ is unitary, gives
\begin{equation}\label{innerprods}
\langle \eta_\lambda(x),\eta_\kappa(x)\rangle=h(x)\langle
\eta'_\lambda(\psi(x)),\eta'_\kappa(\psi(x))\rangle
\end{equation}
so that we get, with $\varphi$ and $\varphi'$ defined by \eqref{mapvarphi},
\begin{equation}\label{compmapvar}
\varphi(x)=h(x)\varphi'(\psi(x))\qqq x\in X\,.
\end{equation}
This shows that $q\circ\varphi=q\circ\varphi'\circ\psi$ and thus that
$(q\circ\varphi)_*\mu=(q\circ\varphi')_*\psi_*\mu$. One has $\mu'=h'\psi_*\mu$
where $h'(x')=h(\psi^{-1}(x'))$ for $x'\in X'$. The maps $q\circ\varphi$ and
$q\circ\varphi'$ are isomorphisms of $X$ and $X'$ with the same subset
$Y\subset \per_m(\Lambda)$. The equality
\begin{equation}\label{kofy}
    k(y)=h((q\circ\varphi)^{-1}(y))=h'((q\circ\varphi')^{-1}(y))\qqq y\in Y
\end{equation}
defines a measurable map $k:Y\to \R_+^*$, and one has
\begin{equation}\label{doublerel}
    (q\circ\varphi')_*\mu'=k(q\circ\varphi)_*\mu\,, \ \ \xi'=k^{-1}\xi
\end{equation}
which shows that the measured section  of \eqref{gencons} is an invariant.
\endproof

\begin{defn} \label{defnrelspec1} We define the {\em relative spectrum}
of $M$ relative to $N$ as  the measured section  of $L$:
\begin{equation}\label{relspec1}
    \Sp_N(M)=((q\circ\varphi)_*(\mu)\,,  \xi(y)=\varphi((q\circ\varphi)^{-1}(y))
\end{equation}
defined up to the gauge action of the unitary group $\cU(N)$.
\end{defn}

Given a measured section  $\nu$ of $L$, one can
consider the following canonically associated data
\begin{itemize}
  \item The Hilbert space $L^2(\nu)$ of $L^2$-sections of $\Ss$ on $\per_m(\Lambda)$
  for the following inner product\footnote{independent of the choice of the
  representative $(\mu,\xi)$ of $\nu$.}
  \begin{equation}\label{innprodsect}
    \langle s,s'\rangle=\int \langle s(y),s'(y)\rangle_{\xi(y)}d\mu(y)
  \end{equation}
  \item The vectors $\kappa_\lambda\in L^2(\nu)$, for $\lambda\in \Lambda$,  such that
   $\kappa_\lambda(g)$
  is the class of $\delta_\lambda$ in $\Ss_g=c_c(\Lambda)/J_g$ where $\delta_\lambda\in c_c(\Lambda)$
  is the delta function.
  \item The action of the algebra of bounded measurable functions on $\per_m(\Lambda)$ by
  multiplication in $L^2(\nu)$.
\end{itemize}

By construction this latter action yields a commutative von Neumann algebra $M_\nu$ of multiplicity $m$
in the Hilbert space $L^2(\nu)$. It is continuous iff the measure on $\per_m(\Lambda)$
associated to $\nu$  is diffuse \ie such that the
measure of any point is zero.

\medskip

\begin{defn} \label{defnunita} A measured section  $\nu$ of $L$ is {\em unitary} iff
the vectors $\kappa_\lambda\in L^2(\nu)$ form an orthonormal basis of $L^2(\nu)$.
\end{defn}

In that case the basis $(\kappa_\lambda)$ gives an action of $N$ in $L^2(\nu)$ as the
von Neumann algebra   of diagonal
operators. We call the corresponding pair of von Neumann algebras $(M_\nu,N)$ acting
in $L^2(\nu)$ the
canonical pair associated to $\nu$.

We can now extend Proposition \ref{lempartunit} to the case at hand of two
von Neumann algebras, a discrete and a continuous one.

\begin{thm} \label{specharinv}
The relative spectrum $\nu=\Sp_N(M)$ is a complete invariant of the relative
position of $M$ relative to $N$. It can be any diffuse  measured section  of the
    $\R_+^*$ bundle $L$ fulfilling the unitarity condition.
\end{thm}

\proof By Lemma \ref{lemind}, the relative spectrum is an invariant of the
relative position of $N$ and $M$. Indeed, given $N$ and $M$ acting in the same
Hilbert space $\cH$ one first constructs a unitary isomorphism $\cH\sim
\ell^2(\Lambda)$, uniquely determined up to the action of the the unitary group
$\cU(N)$. Then one uses Lemma \ref{lemind} to get the measured section  $\nu$ of $L$ which
only depends upon the von Neumann subalgebra $M \subset \cL(\ell^2(\Lambda)) $
and is thus unique up to the gauge action of the unitary group $\cU(N)$. To
show that one obtains a complete invariant, it is enough to show that the
original pair is unitarily equivalent to the canonical pair  $(M_\nu,N)$ acting
in $L^2(\nu)$. We take as above a model of the
inclusion $M\subset \cL(\cH)$ given by the action of $L^\infty(X,\mu)$ by
multiplication in the Hilbert $L^2(X,\mu,S)$ of $L^2$-sections of $S$. The
original pair of von Neumann subalgebras $(M,N)$ of $\cL(\cH)$ is unitarily
equivalent to the pair $(FL^\infty(X,\mu)F^*,\ell^\infty(\Lambda))$ acting in
$\ell^2(\Lambda)$ where $F$ is a unitary isomorphism $F:L^2(X,\mu,S)\to
\ell^2(\Lambda)$. The unitary $F$ gives us the orthonormal basis
$\eta_\lambda=F^*(\epsilon_\lambda)$ and:
\begin{itemize}
  \item The  map $\varphi:X\to \her_m(\Lambda)$ of \eqref{mapvarphi}
   with $q\circ\varphi$ injective\footnote{except on a null set.}.
  \item An isomorphism $V$ of hermitian measurable bundles $V:\varphi^*\Ss\sim
  S$.
  \item A unitary isomorphism $W: L^2(\nu)\to
L^2(X,\mu,S)$.
\end{itemize}
At a point $x\in X$ the isomorphism $V(x)$ is given by
\begin{equation}\label{vmap}
V(x)(\delta_\lambda)=\eta_\lambda(x)\qqq \lambda\in \Lambda\,.
\end{equation}
It is unitary by construction and yields a unitary isomorphism $W: L^2(\nu)\to
L^2(X,\mu,S)$ given by
\begin{equation}\label{wmap}
(W\zeta)(x)=V(x)\zeta(\varphi(x))\qqq \zeta \in L^2(\nu)\,, \ x\in X\,.
\end{equation}
For $f\in L^\infty(X,\mu)$ the operator $W^*fW$ acts in $L^2(\nu)$ as
multiplication by the function $g$ which is arbitrary (say $0$) outside
$Y=q\circ\varphi(X)$ and is such that $g(q\circ \varphi(x))=f(x)$ for $x\in X$.
Thus $W^*L^\infty(X,\mu)W$ is the von Neumann algebra of multiplication by bounded measurable
functions on $\per_m(\Lambda)$. The orthonormal basis $W^*\eta_\lambda$ is the canonical basis of the
sections $\kappa_\lambda$ corresponding to the $\delta_\lambda$, thus the unitary $W^*$ gives the
unitary equivalence of the original pair $(M,N)$ with the model $(M_\nu,N)$ acting
in $L^2(\nu)$, \ie the pair canonically associated to the
measured section  $\nu$ of $L$ given by $\nu=\Sp_N(M)$. This shows that the relative spectrum is a complete
invariant of the relative position of $M$ and $N$.

Conversely, given a diffuse unitary measured section  $\nu$ of $L$, one checks that the relative spectrum
of the pair $(M_\nu,N)$ acting
in $L^2(\nu)$ is equal to $\nu$. \endproof

\subsection{Multiplicity for the discrete algebra}\label{remmult} \hfill \medskip

  The above result extends to the case when the action of
$N$ in $\cH$ is no longer maximal abelian but has some finite multiplicity, \ie we assume
that each minimal idempotent $e_\lambda\in N$ has finite dimensional range $E_\lambda$ in $\cH$.
The quickest way to deal with multiplicity is to
extend $N$ to a maximal abelian algebra $\tilde N\supset N$ and take the invariant
$\Sp_{\tilde N}(M)$ but use instead of the gauge group $\cU(\tilde N)$ the larger group which is the
unitary group $\cU(N')$ of the commutant $N'$ of $N$. The adjoint action of this group on the
space $\her_m(\tilde N)$ of positive hermitian forms $g_{\alpha\beta}$ with $\alpha,\beta\in \Sp(\tilde N)$
is obtained using the natural projection $r:\Sp(\tilde N)\to \Sp(N)$ and identifying $\cU(N')$ with the
group of unitary matrices $u_{\alpha\beta}$ with $\alpha,\beta\in \Sp(\tilde N)$ such that
$u_{\alpha\beta}=0$ except when $r(\alpha)=r(\beta)$. One then has
\begin{equation}\label{Adjact}
    {\rm Ad}(u)(g)=ugu^*\qqq g\in \her_m(\tilde N)\, , \ u\in \cU(N')\,.
\end{equation}
The relative spectrum $\Sp_N(M)$ is then $\Sp_{\tilde N}(M)$ modulo the adjoint action \eqref{Adjact}
of $\cU(N')$.

\smallskip

It is important however to give a more intrinsic definition.
In general one is given for each $\lambda\in \Sp(N)$ a finite dimensional Hilbert space $E_\lambda$
which is the range of the corresponding minimal projection $e_\lambda\in N$. One replaces
$c_c(\Sp(N))$ in \eqref{innprod} and \eqref{radical} by the space $c_c(\Sp(N),E)$ of sections with
finite support of the Hermitian bundle $E$. In the definition of $\her_m(N)$ the $g_{\lambda\mu}$
are no longer scalars but are operators
\begin{equation}\label{glambdamu}
    g_{\lambda\mu}\,:\,  E_\mu\to E_\lambda
\end{equation}
which gives meaning to the expression
\begin{equation}\label{notationinner}
g_{\lambda\mu} \bar a_\lambda b_\mu =\langle a_\lambda,g_{\lambda\mu}(b_\mu)\rangle\qqq a_\lambda
\in E_\lambda,\ b_\mu\in E_\mu\,.
\end{equation}
One defines $\per_m(N)$ as in \eqref{projperspace} \ie as the quotient of $\her_m(N)$ by
the scaling action of $\R^*_+$. Proposition \ref{structproj} holds, with \eqref{metricherm} replaced
by
\begin{equation}\label{metrichermbis}
    \langle \xi,\eta\rangle_y=\sum \langle {\xi}_\mu, g_{\mu\nu}(\sigma(y))\eta_\nu\rangle
    \qqq \xi,\eta\in c_c(\Sp(N),E)
\end{equation}
The hermitian vector bundle  $\Ss(g)$ is defined in the same way, as the quotient of
$c_c(\Sp(N),E)$ by the radical $J_g$. To define the relative invariant $\Sp_N(M)$, one lets
$F:L^2(X,\mu,S)\to \ell^2(\Sp(N),E)$ be a unitary isomorphism. One adapts \eqref{mapvarphi} as follows
\begin{equation}\label{mapvarphimult}
\varphi(x)\in \her_m(N)\,, \ \ \langle \xi,\varphi(x)_{\lambda\kappa}\eta\rangle=\langle
F^*(\xi)(x),F^*(\eta)(x)\rangle\qqq \xi\in E_\lambda,\ \eta\in E_\kappa\,.
\end{equation}
Taking an orthonormal basis $\epsilon_\alpha=\epsilon_{\lambda,i}$ in each $E_\lambda$ one has
$$
\langle
F^*(\epsilon_\alpha)(x),F^*(\epsilon_\beta)(x)\rangle=
\langle \epsilon_{\lambda,i},\varphi(x)_{\lambda\kappa}\epsilon_{\kappa,j}\rangle
$$
so that by Lemma \ref{injecphi} the map $q\circ \varphi$ is injective outside a null set.
Definition \ref{defngen} and Lemma \ref{lemind} are unchanged with $\ell^2(\Sp(N),E)$ instead of
$\ell^2(\Sp(N))$. The only change occurs for the gauge group in   Definition \ref{defnrelspec1}.

\begin{defn} \label{defnrelspec2} We define the {\em relative spectrum}
of $M$ relative to $N$ as  the measured section  of $L$:
\begin{equation}\label{relspec2}
    \Sp_N(M)=((q\circ\varphi)_*(\mu)\,,  \xi(y)=\varphi((q\circ\varphi)^{-1}(y))
\end{equation}
defined up to the gauge action of the unitary group of $\End_N(E)$.
\end{defn}
In the Definition \ref{defnunita} of the unitarity of the measured section  $\nu$ one uses instead
of the vectors $\kappa_\lambda$ the natural linear maps  $\kappa_\lambda:E_\lambda\to L^2(\nu)$ where
for all $\xi \in E_\lambda
    \subset c_c(\Sp(N), E)$ one lets
\begin{equation}\label{kappalambda}
    \kappa_\lambda(\xi)_g={\rm class}\; {\rm of}\; \xi\in c_c(\Sp(N), E)/J_g,\
     \forall g\in \per_m(N) \,.
\end{equation}
Theorem \ref{specharinv} holds unchanged.

\medskip

\section{The unitary (ckm) invariant of Riemannian manifolds} \label{sectchar}

To any compact oriented smooth Riemannian manifold $X$ with Riemannian metric
$g$ we associate the following spectral triple\footnote{where $v$ is the volume form on $X$} $(M,\cH,D)$,
\begin{equation}\label{spectripV}
   M=L^\infty(X,dv)\,, \ \  \cH=L^2(X,\wedge^*)\,,\ \ D=d+d^*\,,
\end{equation}
so that $\cH$ is the Hilbert space of square integrable differential
forms (with complex coefficients) on which
the algebra $L^\infty(X,dv)$ acts by multiplication operators, while $D=d+d^*$ is the
signature operator. In the even dimensional case one uses the volume form (which uses the
orientation)  to define (\cf \cite{lawmich}) the $\Z/2$-grading
$\gamma$ of the Hilbert space $\cH$ while in the odd dimensional case one uses the corresponding
operator $\gamma$ of square\footnote{with suitable powers of $i$} $1$ to reduce the Hilbert space
to the subspace where $\gamma=1$.

\subsection{Completeness of the invariant} \hfill \medskip

Usually a spectral triple is given by restricting to the algebra of smooth functions,
but in the above case the latter algebra can be recovered from it von Neumann algebra weak closure
using the domains of powers of $D$ to define smoothness. More precisely:

\begin{prop}\label{unicity}
The signature spectral triple $(M,\cH,D)$ uniquely determines the compact
 smooth Riemannian manifold $X$.
\end{prop}

\proof Let $X_j$ be two compact oriented smooth Riemannian manifold and $(M_j,\cH_j,D_j)$
the associated triples as in \eqref{spectripV}. Let $U:\cH_1\to \cH_2$ be a unitary operator
such that
\begin{equation}\label{uconjugates}
    UM_1U^*=M_2\,,\ \ UD_1U^*=D_2
\end{equation}
Let $\cH_j^\infty=\cap_n\Dom D_j^n$ be the intersection of domains of powers of the self-adjoint
unbounded operator $D_j$. Then the algebra $C^\infty(X_j)$ is the subalgebra of $M_j=L^\infty(X_j,dv_j)$
given by
\begin{equation}\label{smoothalg}
    \cA_j=\{f\in M_j\,|\, f\cH_j^\infty\subset \cH_j^\infty\}
\end{equation}
as can be seen by applying $f$ to the constant $0$-form $1$ to get the inclusion
$\cA_j\subset C^\infty(X_j)$
while the other inclusion follows since elements of $\cH_j^\infty$ are exactly the smooth forms.
Thus one gets from \eqref{uconjugates} that
\begin{equation}\label{uconjugatesmore}
    UC^\infty(X_1)U^*=C^\infty(X_2)\,,
\end{equation}
and there exists a diffeomorphism $\psi:X_2\to X_1$, such that
\begin{equation}\label{diffeoact}
    UfU^*=f\circ \psi  \qqq f\in C^\infty(X_1)\,.
\end{equation}
The Riemannian metric $g$ is uniquely determined for instance by the equality
\begin{equation}\label{metricg}
    [D,f]^2\in \cA\,, \ [D,f]^2=-g^{\mu\nu}\partial_\mu f\partial_\nu f \qqq f=f^*\in \cA
\end{equation}
\endproof

\begin{cor} \label{coruniqueness} The pair given by the
spectrum\footnote{with multiplicities} of $D$ and the relative spectrum
$\Sp_N(M)$, where $N$ is the von Neumann algebra of functions of $D$, uniquely
determines the compact smooth Riemannian manifold $X$.
\end{cor}

\proof The knowledge of the spectrum of $D$ with the multiplicities gives the operator
$D$ acting in the Hilbert space $\cH$ and the knowledge of the relative spectrum
$\Sp_N(M)$ gives, by Theorem \ref{specharinv}, the
von Neumann subalgebra $M\subset \cL(\cH)$ and hence the triple $(M,\cH,D)$ of Proposition \ref{unicity}.
\endproof

\begin{rem}\label{laplacian} {\rm Proposition \ref{unicity} has an analogue where
one uses the scalar Laplacian $\Delta$ instead of the signature operator. To determine the
metric one uses
\begin{equation}\label{metricglap}
    [[\Delta,f],g]\in \cA\,, \ [[\Delta,f],g]=-2g^{\mu\nu}\partial_\mu f\partial_\nu g \qqq f=f^*,g=g^*\in \cA
\end{equation}
In fact one can also use the embedding results of \cite{bbg} which show how to recover
the metric at the local level from the heat expansion. One definite advantage in
using the Laplacian is that the eigenfunctions can be chosen to be real valued which
further reduces the gauge group to the unitary group of the self-adjoint real subalgebra
$N_{\rm sa}=\{ x\in N\,|\, x=x^*\}$.
Order one operators are however easier to characterize than order two operators, in particular for the
orientability condition as we shall see below in \S \ref{chararange}.
}\end{rem}

\medskip

\subsection{The spectral meaning of points}\hfill \medskip \label{spectralpoints}

Once we know the spectrum\footnote{with multiplicities} $\Lambda$ of $D$,   the
missing information contained in $\Sp_N(M)$ is a measured section  $\nu$ of the $\R_+^*$ bundle $L$ over $\per_m(\Lambda)$
fulfilling the unitarity condition. It should be interpreted as giving the probability for correlations
between the possible frequencies, while a ``point" of the geometric space $X$ can be
thought of as a correlation, \ie a specific positive  hermitian matrix $g_{\lambda\kappa}$ (up to scale) in the
support of $\nu$. To go further in this ``spectral"  identification of points one needs
to check the injectivity of the map $\varphi$ of \eqref{mapvarphi} at the topological rather
than at the measure theoretic level of Lemma \ref{injecphi}.

\begin{lem}\label{injectsmooth} Let $X$ be a compact oriented smooth Riemannian manifold
and $(M,\cH,D)$ the spectral triple of \eqref{spectripV}. Then the map $\varphi$ of
\eqref{mapvarphi} is injective.
\end{lem}

\proof By construction one has
\begin{equation}\label{mapvarphismooth}
\varphi(x)\in \her_m(N)\,, \ \ \varphi(x)_{\lambda\kappa}=\langle
\eta_\lambda(x),\eta_\kappa(x)\rangle
\end{equation}
where the $\eta_\lambda$ form an orthonormal basis of eigenfunctions for $D$. The
space $C^\infty(X,S)$ of smooth sections of the hermitian vector bundle $S$ over $X$
on which $D$ is acting coincides with the intersection of the domains of powers of
$D$ and hence with the following vector space
\begin{equation}\label{smoothsect}
   C^\infty(X,S)=\{\sum a_\lambda\eta_\lambda\,|\, a\in \cS\}
\end{equation}
where $\cS$ is the Schwartz space of sequences $(a_\lambda)$ of rapid decay. For any pair
$\alpha=\sum a_\lambda\eta_\lambda$, $\beta=\sum b_\lambda\eta_\lambda$
of elements of  $C^\infty(X,S)$, one has
$$
\langle
\alpha(x),\beta(x)\rangle=\sum \bar a_\lambda b_\kappa
\varphi(x)_{\lambda\kappa}\qqq x\in X \,,
$$
where convergence can be checked using Sobolev estimates.
Since any smooth function $f\in C^\infty(X)$ can be written in the form
$f(x)=\langle
\alpha(x),\beta(x)\rangle$ one gets the required injectivity of $\varphi$.
\endproof

\medskip

\subsection{The characterization of the range}\hfill \medskip \label{chararange}

The really difficult problem, then, is to characterize which values of these invariants
correspond to compact smooth Riemannian manifolds. We shall only deal below with the case of spin$^c$
manifolds in which case we use the Dirac spectral triple instead of the above signature triple.

The key result that we shall us is that under the simple \axioms of
\cite{CoSM} on a spectral triple $(\cA,\cH,D)$, with  $\cA$ {\em
commutative},  the algebra $\cA$ is the algebra $C^\infty(X)$ of
smooth functions on a (unique) smooth compact manifold $X$. The five
\axioms (\cite{CoSM}), in dimension $p$, are

\begin{enumerate}
  \item  The  $n$-th characteristic value of the resolvent of $D$ is
$O(n^{-1/p})$.
  \item $\left[ [D,a],b\right] = 0 \qquad  \forall \,
a,b \in \cA$.
  \item  For any $a\in \cA$ both $a$
and $[D,a]$ belong to the domain of $\delta^m$, for any integer $m$
where $\delta$ is the derivation: $\delta(T)=[|D|,T]$.
  \item  There exists $c\in\cA^{\otimes n}$, $n=p+1$, totally antisymmetric in its last $p$-entries, and
such that\footnote{This assumes $p$ odd, in  the even case one requires that
 $\pi_D(c)=\gamma$ fulfills $\gamma = \gamma^* ,  \gamma^2 = 1, \gamma D
=-D\gamma$.}
\begin{equation}\label{second}
\pi_D(c) =1\,, \ {\rm where} \ \  \pi_D(a_0\otimes \cdots \otimes
a_p)=a_0[D,a_1]\cdots [D,a_p]\qqq a_j\in\cA\,.
\end{equation}
  \item Viewed as an $\cA$-module the space $\cH_{ \infty} =
  \cap \Dom D^m$ is
finite and projective. Moreover the following equality defines a
hermitian structure $( \ |\ )$ on this module: $
    \langle  \xi ,a \,\eta  \rangle = \cutint \, a (\xi |\eta) \, |D|^{-p}
\,, \ \forall   a \in \cA   ,   \forall  \xi ,\eta \in \cH_{ \infty}
$.
\end{enumerate}
In the last equation, $\cutint$ is the noncommutative integral given by the Dixmier trace.

\medskip

We can now restate  Theorem 11.5 of \cite{CoRec} as:

\begin{thm} \label{compdisccont} Let $(\cA,\cH,D)$ fulfill the above five
conditions  and assume that the multiplicity is $m=2^{p/2}$, then there exists a unique smooth compact
oriented spin$^c$ Riemannian manifold $(X,g)$ such that the triple $(\cA,\cH,D)$
is given by
\begin{itemize}
  \item $\cA=C^\infty(X)$.
  \item $\cH=L^2(X,S)$ where $S$ is the spinor bundle.
  \item $D$ is a Dirac operator associated to the Riemannian metric $g$.
\end{itemize}
\end{thm}

 Note that there is no uniqueness of $D$ since we only know its
principal symbol. This is discussed in \cite{CoSM} and \cite{FGV}. The Hochschild cycle
$c$ gives the orientation.

Let us start with the spectrum $\Lambda=\Sp(D)$ given as a subset of $\R$ with
multiplicity and \axiom (1) determines the growth of $\Lambda$. This fixes the Hilbert space $\cH=\ell^2(\Lambda)$ and the
self-adjoint operator $D$ which is just a multiplication operator. The analogue
in our context of the geodesic flow is the following one parameter group
\begin{equation}\label{geodesicflowdefn}
\gamma_t(T)=e^{it|D|} Te^{-it|D|} \qqq T\in \cL(\cH)\,.
\end{equation}

\begin{defn}\label{classcinfty}
We say  that an operator $T\in \cL(\cH)$ is of class $C^\infty$ when the map
from $\R$ to $ \cL(\cH)$ given by $t\mapsto \gamma_t(T)$ is of class $C^\infty$
(for the norm topology of $\cL(\cH)$) and we denote by $C^\infty(\cH,D)$ this
subalgebra of $\cL(\cH)$.
\end{defn}

This algebra only depends upon $(\cH,D)$.

\begin{defn} \label{defnunitasmooth} A unitary measured section  $\nu$ of $L$ is {\em smooth} when
\begin{itemize}
  \item The support $K$ of $\nu$ is compact.
  \item The map $a\in \cS(\Lambda)\mapsto \sum a_\lambda\kappa_\lambda$ is an isomorphism
  of the Schwartz space $\cS(\Lambda)$ with $C^\infty(K,\Ss)$.
  \item Any element of $C^\infty(K)$ is of class $C^\infty$ in the sense of Definition \ref{classcinfty}.
\end{itemize}
\end{defn}

\medskip
Let then  $\nu$ be a smooth unitary measured section  of $L$. The order one \axiom (2) means that $D$ is a differential operator
of order one. The regularity \axiom (3) is now
automatically fulfilled. The orientability \axiom (4) can be formulated as the
existence of a $p$-form $c$ on $K$ such that \eqref{second} holds. Finally \axiom (5) now
becomes an equation.

\medskip

\section{The  Sunada examples}\label{sectexamples}

We have not gone very far in computing examples. The case of flat tori is straightforward,
the only difficulty being to properly take care of the gauge ambiguity coming from the
multiplicity of eigenvalues. Thus, instead, we shall concentrate on the examples of isospectral
non-isometric Riemannian manifolds constructed by Sunada in \cite{Sunada}.
Let, as in \cite{Sunada}, $G$ be a  finite group and $H_1$, $H_2$ be subgroups of $G$ such that
each conjugacy class $[g]\in [G]$ meets $H_1$ and $H_2$ in the same number of elements.
The existence of examples where the $H_j$ are non-isomorphic was used in \cite{Sunada} to
produce classes of examples of isospectral but non-isomorphic Riemannian geometries.
One considers a compact oriented smooth Riemannian manifold $X$ on which the
group $G$ acts freely by isometries. One then takes the quotient Riemannian manifolds
$X_j=H_j\backslash X$. For each $y\in Y=G\backslash X$ the fiber $p^{-1}(y)$ of the
projection $p:X\to Y$ can be identified with $G$ as a $G$-space for the left action
of $G$  but there is a non-canonical choice of base point in the fiber. Two different choices
are related by the right action of $G$. The essence
of the situation is captured by the comparison, commuting with
 the right representation $\rho$ of $G$, of the von Neumann algebras
$M_j$   of multiplication by functions on $H_j\backslash G$
in the Hilbert space $\ell^2(H_j\backslash G)$.
The algebra that plays the role of the
algebra of functions of $D$ is a  subalgebra of the center of the convolution algebra $C^*(G)$.
It acts on $\ell^2(H_j\backslash G)$ by right
convolution which commutes with the projections
$P_j$: $\ell^2(G)\to \ell^2(H_j\backslash G)$ given by averaging
over $H_j$ \ie
$$
P_j(\xi)(g)= (\# H_j)^{-1}\sum_{H_j}\xi(hg)
$$
By Lemma 2 of \cite{Sunada} the representations $\pi_j$ by right convolution of $C^*(G)$ in the Hilbert spaces
$\cH_j=\ell^2(H_j\backslash G)$ are equivalent since their characters are the same. Moreover this
character is given by
\begin{equation}\label{charofres}
    \tr(\pi_j(g))=\tr(\rho(g)P_j)=(\# H_j)^{-1}\sum_{[k]\in [G]}\#([k]\cap H_j)\tr(\lambda(k)\rho(g))
\end{equation}
where $\lambda$ is the left regular representation of $G$ and $\rho$ the right regular one. On the canonical
basis $\epsilon_h$, $h\in G$ of $\ell^2(G)$, the operator $ \lambda(k)\rho(g)$ is a permutation
and its trace is the number of fixed points, \ie the cardinality of
$\{x\in G\,|\, kxg^{-1}=x\}$. It is non-zero
when the conjugacy class of $k$ is the same as that of $g$ and is equal, in that
case, to the order
of the centralizer $C_g$. The latter is $\# C_g=\# G/\# [g]$ so that one gets
\begin{equation}\label{charofres1}
    \tr(\pi_j(g))=(\# G)(\# H_j)^{-1}(\#([g]\cap H_j))(\# [g])^{-1}
\end{equation}
The two projections $P_j$ are equivalent in the von Neumann algebra $\lambda(G)$ of the
left-regular representation of $G$. One looks in fact for a unitary $U$ such that
\begin{equation}\label{unitequiv}
   U\in \lambda(G)\,,\ \  UU^*=U^*U=1\,,\ \ UP_1U^*=P_2\,, \ \ UM_1U^*=M_2
\end{equation}
where $M_j$ is the algebra of multiplication by functions on $H_j\backslash G$. Since
the two projections $P_j$ are equivalent in $\lambda(G)$ one can find a
unitary $U_0\in \lambda(G)$ which fulfills all but the last conditions in \eqref{unitequiv}.
The remaining freedom is a unitary $V$ in the reduced algebra $P_1\lambda(G)P_1$. One looks for
$V$ such that
\begin{equation}\label{unitequivbis}
   V\in P_1\lambda(G)P_1\,,\ \  VV^*=V^*V=P_1\,, \ \ VM_1V^*=U_0^*M_2U_0\,.
\end{equation}
We let $N$ be the quotient of $C^*(G)$ by the kernel of the representation $\pi_j$ and identify
it with its image under $\pi_j$. Then we are comparing the two pairs $(M_j,N)$ as in Proposition \ref{lempartunit}
with $N$ no longer commutative.

\medskip

\begin{prop}\label{conjugsubgroups} The two pairs $(M_j,N)$ are conjugate iff the subgroups
$H_j\subset G$ are conjugate.
\end{prop}

\proof The data $(M_j,N)$ with $N=C^*(G)/\ker \pi_j$, gives the set $\Sp(M_j)$ together with the
action of $G$ on this set. Thus it gives back the $G$-space $H_j\backslash G$. This in turns
determines uniquely the conjugacy class of the subgroup $H_j\subset G$ from the
isotropy group of any point. Conversely an isomorphism
of the $G$-spaces $H_j\backslash G$ yields an isomorphism of the $(M_j,N)$.
\endproof

More generally, one can determine the behavior of the relative spectrum under the
action of a finite group $H$ of automorphisms. We let $H$ be a finite group
unitarily represented in the Hilbert space $\cH$
of a spectral triple $(M,\cH,D)$ so that
\begin{equation}\label{isoconds}
    \pi(h)D=D\pi(h)\,, \ \ \pi(h)M\pi(h)^*=M\qqq h\in H\,.
\end{equation}
The spectral triple $(M_H,\cH_H,D_H)$ is obtained as  follows
\begin{equation}\label{fixedspec}
    M_H=M\cap \pi(H)'\,, \ \cH_H=\{\xi\in \cH\,|\, \pi(h)\xi=\xi ,\forall h\in H\}\,, \ D_H=D|_{\cH_H}\,.
\end{equation}
so that the von Neumann algebra $M_H$ is the fixed point algebra under the action of $H$,
$$
M_H=\{x\in M\,|\, \pi(h)x\pi(h)^*=x,\forall h\in H\}\,.
$$
For each $\lambda\in \Sp(D)$ one lets $E_\lambda\subset \cH$ be the corresponding eigenspace. Then let
$\pi_\lambda$ be the restriction of the
representation $\pi$ to $E_\lambda$. The eigenspace $F_\lambda$ of the spectral triple
$(M_H,\cH_H,D_H)$ is obtained as  follows
\begin{equation}\label{eigenfixed}
   F_\lambda=E_\lambda^H=\{\xi\in E_\lambda\,|\, \pi(h)\xi=\xi ,\forall h\in H\}
\end{equation}
We  denote by $P_\lambda$ the orthogonal projection from $E_\lambda$ to $F_\lambda$.
In order to compute the relative spectrum of $M_H$ relative to the algebra $N_H$ of functions of
$D_H$ we use the  intrinsic formulation of \S \ref{remmult} on multiplicity.

\begin{lem} \label{lemofproj} Assume that the action of
$H$ on $M$ is free (\ie the set of fixed points $x\in X, \exists h\neq 1, hx=x$ is
negligible). The following defines a map from $\Sp_N(M)\subset\her_m(N)$ to $\her_m(N_H)$
\begin{equation}\label{sectmap}
    p(g)_{\lambda\kappa}=P_\lambda g_{\lambda\kappa}P_\kappa
\end{equation}
The map $p$ is invariant under the adjoint action of $H$ and $\Sp_{N_H}(M_H)$ is the image of $\Sp_N(M)$
under $p$.
\end{lem}

We leave the proof as an exercise.
We shall now show  an explicit example where the two pairs $(M_j,N)$ of Proposition \ref{conjugsubgroups}
are not conjugate but become so up to an automorphism of $N$.
We use a concrete example of triple $(G,H_1,H_2)$ obtained as follows (\cf \cite{Sunada}). One takes
for $G$ the semi-direct product of the additive group $\Z/8\Z$ by the action of the multiplicative
group $(\Z/8\Z)^*=\{1,3,5,7\}$. It is a group of order $32$ with multiplication given by
\begin{equation}\label{multilaw}
(a,b).(c,d)=(ac,b+ad) \in (\Z/8\Z)^*\times \Z/8\Z
\end{equation}
The adjoint action ${\rm Ad}(g)$ of $G$ on itself is given by
$$
{\rm Ad}(g)(x,y)=(x, a y + (1-x) b) \qqq g=(a,b)\in G
$$
The conjugacy classes $[g]\in [G]$ are the following
$$
\{(1,0)\}\,,\{(1,4)\}\,,\{(1,2),(1,6)\}\,,\{(1,1),(1,3),(1,5),(1,7)\}\,,
$$
$$
\{(3,0),(3,2),(3,4),(3,6)\}\,,\{(3,1),(3,3),(3,5),(3,7)\}
$$
$$
\{(5,0),(5,4)\}\,,\{(5,2),(5,6)\}\,,\{(5,1),(5,3),(5,5),(5,7)\}
$$
$$
\{(7,0),(7,2),(7,4),(7,6)\}\,,\{(7,1),(7,3),(7,5),(7,7)\}
$$
The subgroup $H_1=\{(a,0)\,|\,a\in (\Z/8\Z)^*\}$ has the same cardinality of
intersection with the conjugacy classes of $G$ as the subgroup
$$
H_2=\{(1,0),(3,4),(5,4),(7,0)\}
$$
The character of the representation of $G$ given by \eqref{charofres1} is the central
function which vanishes except on the $4$ conjugacy classes which meet the $H_j$.
The characters of irreducible representations of $G$ are given from the orbits of the
action of  $(\Z/8\Z)^*$ on the dual of $\Z/8\Z$ which we identify with $\Z/8\Z$ using the
basic character $n\in \Z/8\Z\mapsto \chi(n)=\omega^n$ where $\omega=e^{i\pi/4}$ is
a primitive $8$-th root of $1$.
Thus each $n\in \Z/8\Z$ gives the character $m\mapsto \chi(nm)$ of $\Z/8\Z$. The eleven characters of $G$
are\footnote{With the label of the orbit of  $(\Z/8\Z)^*$ on  $\Z/8\Z$.}

\begin{enumerate}
  \item[$0$] Four $1$-dimensional characters
  $(a,b)\mapsto \kappa(a)$ where $\kappa$ is a character of $(\Z/8\Z)^*$.
  \item[$4$] Four $1$-dimensional
  characters $(a,b)\mapsto \kappa(a)\chi(4b)$ where $\kappa$ is a character of $(\Z/8\Z)^*$.
  \item[$2$] Two $2$-dimensional characters given by
   \begin{equation}\label{2dchar}
   c_2((a,b))= \begin{cases} 0~&\text{if}\quad a\notin \{ 1,5\} \ \text{or} \ b\notin 2\Z
  \\
(-1)^{b/2}2~&\text{if}\ a \in \{ 1,5\}, \ b\in 2\Z.\end{cases}
  \end{equation}
   \begin{equation}\label{2dcharbis}
   c'_2((a,b))= \begin{cases} 0~&\text{if}\quad a\notin \{ 1,5\} \ \text{or} \ b\notin 2\Z
  \\
(-1)^{b/2+(a-1)/4}2~&\text{if}\ a \in \{ 1,5\}, \ b\in 2\Z.\end{cases}
  \end{equation}
  \item[$1$] One $4$-dimensional character given by
  \begin{equation}\label{4dchar}
   c_4((a,b))= \begin{cases} 0~&\text{if}\quad a\neq 1 \ \text{or} \ b\notin 4\Z
  \\
(-1)^{b/4}4~&\text{if}\ a=1, \ b\in 4\Z.\end{cases}
  \end{equation}
\end{enumerate}
The idempotent\footnote{The Haar measure on $G$ is normalized of total mass $1$}  $4c_4$ is  the sum of the following $4$ minimal projections $e_m\in C^*(G)$, for $m\in
\{1,3,5,7\}$,
\begin{equation}\label{ejproj}
     e_m((a,b))= \begin{cases} 0~&\text{if}\quad a\neq 1
  \\
4\chi(mb)~&\text{if}\ a =1.\end{cases}
\end{equation}

The $8$-dimensional representation $\pi_j$ of $G$ in $\ell^2(H_j\backslash G)$
decomposes as the direct sum of $4$ irreducible
representations, each of multiplicity one, corresponding to the trivial character, the character
$(a,b)\mapsto \chi(4b)$ and the characters $c_2$ and $c_4$. We now identify the coset spaces
$H_j\backslash G$ with $\Z/8\Z$, just as sets, to keep track of the algebras $M_j$ of
multiplication operators. It is simpler for the right coset spaces $G/H_j$. By \eqref{multilaw}
the map $p_1(a,b)=b$ gives a bijection $G/H_1\to \Z/8\Z$. Similarly
the bijection $p_2:G/H_2\to \Z/8\Z$ is obtained by intersecting each class with the subgroup
\begin{equation}\label{subgroup8}
    L=\{(1,b)\,|\, b\in \Z/8\Z\}\subset G
\end{equation}
which is transverse to the subgroups $H_j$. We use the identification $H_j\backslash G\sim L$ for
the left coset spaces and thus obtain two representations $\pi_j$ of $G$ in $\ell^2(L)$.

\begin{lem}\label{explexample} The von Neumann algebras $\pi_j(G)''$ are the same and the
representations $\pi_j$ are inner related as follows
\begin{equation}\label{innerrelreps}
    \pi_2(g)=\pi_1(UgU^*)\qqq g\in G
\end{equation}
where $U\in C^*(G)$ is the unitary
\begin{equation}\label{unitaryU}
    U=1-2(e_3+e_5)
\end{equation}
\end{lem}

\proof The character of the representation $\pi_j$ of $G$ in $\ell^2(H_j\backslash G)$
is the central function $\tau(g)$ which assigns to $g\in G$ the number of fixed points of
right translation by $g$ in $H_j\backslash G$, it is given by
\eqref{charofres1}. Thus the inner product of $\tau$ with the character $\alpha$
 of a representation
of $G$ is proportional to $\sum_{H_j}\alpha(g)$. Thus to know which representations
of $G$ appear in $\pi_j$, one just computes $\sum_{H_j}\alpha(g)$ for the irreducible
characters $\alpha$ listed above. It is enough to do it for $H_1=(\Z/8\Z)^*$. One gets
a non-zero result only for the trivial character, the character
$(a,b)\mapsto \chi(4b)$ and the characters $c_2$ and $c_4$. Since the dimension of
$\ell^2(H_j\backslash G)$ is $8$ the multiplicities are all equal to $1$. We decompose the
idempotent $2c_2$ as the sum of the following $2$ minimal projections $e_m\in C^*(G)$, for $m\in
\{2,6\}$,
\begin{equation}\label{ejproj1}
     e_m((a,b))= \begin{cases} 0~&\text{if}\quad a\notin \{1,5\}
  \\
2\chi(mb)~&\text{if}\ a \in \{1,5\}.\end{cases}
\end{equation}
We extend the notation for  $m\in
\{0,4\}$ using the characters of one dimensional irreducible representations
\begin{equation}\label{ejproj2}
e_m((a,b))=\chi(mb)\qqq m\in
\{0,4\}\,.
\end{equation}
By construction the $\pi_j(e_m)$ are the minimal
projections of a maximal abelian algebra acting in $\ell^2(H_j\backslash G)$. One checks that
under the isomorphism $\ell^2(H_1\backslash G)\sim \ell^2(L)\sim \ell^2(H_2\backslash G)$
one gets $\pi_1(e_m)=\pi_2(e_m)$ for all $m$. In fact it is enough to show that with
$\sigma=\sum \chi(m)e_m$ the operator $\pi_j(\sigma)$ is the translation of $1$ in $\ell^2(L)$.
One finds (with Haar measure normalized as a probability measure) that
\begin{equation}\label{sigma}
    (\pi_j(\sigma)f)(b)=f(b+1)\qqq f\in \ell^2(L)\,, \ b\in L\sim \Z/8\Z\,.
\end{equation}
It follows from the equality $\pi_1(e_m)=\pi_2(e_m)$ for all $m$ that the
von Neumann algebras $\pi_j(G)''$ are the same, since the partial isometry
realizing the equivalence of minimal projections is unique up to a phase. One checks
\eqref{innerrelreps} by direct calculation.
\endproof

We let the two pairs $(M_j,N)$ be as in Proposition \ref{conjugsubgroups}.

\begin{prop}\label{moduloauto} Let $G$ and $H_j\subset G$ be as above then
the following two pairs of von Neumann algebras are conjugate
\begin{equation}\label{pairsofvon}
    (M_1,\pi_1(G)'')\sim (M_2,\pi_2(G)'')\,, \ M_j=\ell^\infty(H_j\backslash G)
\end{equation}
Any  equivalence \eqref{pairsofvon} induces a non-trivial automorphism
of $N=C^*(G)/\ker \,\pi_j$.
\end{prop}

\proof The isomorphism $\ell^2(H_1\backslash G)\sim \ell^2(L)\sim \ell^2(H_2\backslash G)$
transforms $M_1$ into $M_2$ and preserves the  von Neumann algebras $\pi_j(G)''$
thus it induces the equivalence \eqref{pairsofvon}. This proves the first
statement. To get the second it is enough to show that one cannot find a
 unitary $V$ acting in $\ell^2(H_1\backslash G)$ and such that
$$
V\ell^\infty(H_1\backslash G)V^*=\ell^\infty(H_1\backslash G)\,, \ \
 V\pi_1(g)V^*=\pi_1(UgU^*)\qqq g\in G
$$
It follows that $V=\pi_1(U)Z$ where $Z$ is a unitary in the commutant
of $\pi_1(G)$. This commutant is the following algebra
$$
C=\{ \lambda_0e_0+ \lambda_4e_4 + 2 \lambda c_2 + 4\mu c_4\,|\, \lambda_j,\lambda,\mu \in \C   \}
$$
Now the normalizer of $M_1=\ell^\infty(H_1\backslash G)$ intersected with the
maximal abelian algebra generated by the $e_m$ is, up to a scalar factor of modulus one,
the group of order $8$ of translations generated by $\pi_1(\sigma)$.
Thus one checks that for no unitary element $Z$ the product $V=\pi_1(U)Z$ is in the
normalizer of $M_1=\ell^\infty(H_1\backslash G)$.
\endproof


\begin{thebibliography}{99}


\bibitem{Autonne}
L.~Autonne,  {\em Sur les matrices hypohermitiennes et les unitaires}, Comptes Rendus de
l'Acad\'emie des Sciences, Paris, vol. 156 (1913),   858--860.


\bibitem{bbg}
P.~B\'erard, G.~Besson, S.~Gallot {\em Embedding Riemannian manifolds by their heat kernel},
Geometric And Functional Analysis, Volume 4, Number 4 (1994), 373--398.

\bibitem{Browne}
E. T. Browne, {\em The characteritic roots of a matrix},
Bull. Amer. Math. Soc.
vol. 36 (1930),  705--710.

\bibitem{cc2} A.~Chamseddine, A.~Connes, {\em  The Spectral
 action principle}, Comm. Math. Phys. Vol.186 (1997), 731--750.




\bibitem{mc2} A.~Chamseddine, A.~Connes, M.~Marcolli,
{\em Gravity and the standard model with neutrino mixing}, hep-th/0610241.







\bibitem{Co-book}  A.~Connes, {\it Noncommutative geometry},
 Academic Press (1994).

\bibitem{Co-spec}  A.~Connes, {\em Geometry from the spectral point of view}, Lett. Math.
Phys. 34 (1995), 203–238

\bibitem{CoSM} A.~Connes, {\em Gravity coupled with matter and the
foundation of noncommutative geometry}, Comm. Math. Phys. (1995)

\bibitem{cmindex}  A. Connes, H. Moscovici, {\em
 The local index formula in noncommutative
geometry},  GAFA, Vol. 5 (1995), 174--243.


\bibitem{CoRec} A.~Connes, {\em On the spectral characterization of manifolds},
arXiv Math.OA 0810.2088.


\bibitem{dix} J. Dixmier, {\it Les C*-alg\`ebres et leurs representations}
Reprint of the second (1969) edition. Les Grands Classiques
Gauthier-Villars. \'Editions Jacques Gabay, Paris, 1996. 403 pp.

\bibitem{EY} C.~Eckart, G.~Young,
{\em A principal axis transformation for non-hermitian matrices},
 Bull. Amer. Math. Soc. 45, no. 2 (1939),  118--121.


 \bibitem{FGV} H.~Figueroa, J.M.~Gracia-Bond\'ia, J.~Varilly, {\em
Elements of Noncommutative Geometry}, Birkh\"auser, 2000.

\bibitem{Haagerup} U.~Haagerup,
{\em  Orthogonal maximal abelian $*$-subalgebras of the $n\times n$ matrices and cyclic $n$-roots}.
Operator algebras and quantum field theory (Rome, 1996), 296--322, Int. Press, Cambridge, MA, 1997.

\bibitem{lawmich} H. B.~Lawson, M-L.~Michelsohn {\em Spin geometry},
Princeton Mathematical Series, 38. Princeton University Press,
Princeton, NJ, 1989.

\bibitem{Milnor} J.~Milnor, {\em Eigenvalues of the Laplace operator
on certain manifolds} Proc. Natl. Acad. Sci. U S A.   51(4) (1964), 542.

\bibitem{Nico} R.~Nicoara,  {\em A finiteness result for commuting squares of matrix algebras}. J. Operator Theory 55 (2006), no. 2, 295--310

\bibitem{Pedersen} G.~Pedersen, {\em $C^*$-algebras and their Automorphism groups} London Math. Society
monographs 14. Academic Press (1979).

\bibitem{Popa} S.~Popa, {\em Orthogonal pairs of $*$-subalgebras in finite von Neumann algebras}.
J. Operator Theory 9 (1983), no. 2, 253--268.

\bibitem{ReVa} A.~Rennie, J.C.~Varilly, {\em Reconstruction of
manifolds in noncommutative geometry}, math.OA/0610418.

\bibitem{vNeumann} J.~von Neumann, {\em Zur Algebra
der Funktionaloperatoren und Theorie der normalen Operatoren},
Math. Ann. 102 (1929) 370-427.

\bibitem{vNeumann1} J.~von Neumann, {\em On rings of operators: reduction theory},
Ann. Math. 50 (1949) 401-485.


\bibitem{Sunada} T.~Sunada,
{\em  Riemannian coverings and isospectral manifolds}.
Ann. of Math. (2) 121, no. 1, (1985),  169--186.

\bibitem{Sylvester} J.~Sylvester, {\em On the reduction of a bilinear
quantic of the $n$'th order to the form of a sum of
$n$ products by a double orthogonal substitution}, Messenger of Mathematics, vol. 19 (1889), 42--46.

\bibitem{Taber} H.~Taber {\em On the linear transformations
between two quadrics}, Proceedings of the London Mathematical
Society, vol. 24 (1892-93),   290--306.
\end{thebibliography}
\end{document}